\documentclass{aa}
\usepackage{graphicx}
\usepackage{txfonts}
\usepackage{natbib}
\usepackage{adjmulticol}
\usepackage{lipsum}
\bibpunct{(}{)}{;}{a}{}{,}
\usepackage[usenames]{color}
\usepackage{soul}

\begin{document} 

\title{Modeling the Galactic center gamma-ray emission with more realistic cosmic-ray dynamics}
\subtitle{}

\author{Andrés Scherer\inst{1}
\and Jorge Cuadra\inst{2}
\and Franz E. Bauer\inst{1,3,4}}

\institute{Instituto de Astrofísica, Pontificia Universidad Católica de Chile, Av. Vicuña Mackenna 4860, 7820436 Macul, Santiago, Chile
\and Departamento de Ciencias, Facultad de Artes Liberales, Universidad Adolfo Ibáñez, Av. Padre Hurtado 750, Viña del Mar, Chile
\and Millenium Institute of Astrophysics, Vicuña Mackenna 4860, 7820436 Macul, Santiago, Chile
\and Space Science Institute, 4750 Walnut Street, Suite 2015, Boulder, CO 80301, USA}

\date{Received date / Accepted date }

\abstract{Very-high-energy gamma-ray observations of the Galactic center (GC) show extended emission that is strongly correlated
with the morphology of the central molecular zone (CMZ). The best explanation for that emission is a hadronic interaction between cosmic rays (CRs) and ambient gas, where a CR central and continuous source accelerates protons up to 1 PeV ("PeVatron"). However, current models assume very simplistic CR dynamics.}
{Our goal is to verify if more realistic CR dynamics for the GC environment are consistent with current gamma-ray observations, and whether they could be constrained by upcoming observations with the Cherenkov Telescope Array (CTA).}
{We generated synthetic gamma-ray maps using a CR transport model with spherical injection, different diffusion regimes (in and out of the CMZ), polar advection, and mono-energetic particles of 1 PeV, and including different CR populations injected from the Arches, Quintuplet, and nuclear clusters of young massive stars, plus supernova Sgr A East. We adopted two different 3D gas distributions consistent with the observed gas column density, either with or without an inner cavity.}
{In order to reproduce the existing observations detected by the High Energy Stereoscopic System (HESS), a ring-like gas distribution, with its mass set by the standard Galactic CO-to-H$_2$ conversion factor, and CR acceleration from all relevant sources are required. For a conversion factor one order of magnitude lower, injection rates that are ten times higher are needed. We show that CTA will be able to differentiate between models with different CR dynamics, proton sources, and CMZ morphologies, owing to its unprecedented sensitivity and angular resolution.}
{More realistic CR dynamics suggest that the CMZ has a large inner cavity and that the GC PeVatron is a composite CR population accelerated by the Arches, Quintuplet, and nuclear star clusters, and Sgr A East.} 


\maketitle

\section{Introduction}

There is growing evidence that the Galactic center (GC) plays an active role in the acceleration of cosmic rays (CRs). Very-high-energy gamma-ray observations (0.1--100 TeV) of the central molecular zone (CMZ) by the High Energy Stereoscopic System \citep[HESS;][]{Aharonian.2004b,Aharonian.2009,HESS.2016,HESS.2018}, the Major Atmospheric Gamma-Ray Imaging Cherenkov \citep[MAGIC;][]{MAGIC.2020}, and the Very Energetic Radiation Imaging Telescope Array System \citep[VERITAS;][]{VERITAS.2021} all support a particle acceleration scenario. In these observations, extended emission correlates with the gas morphology in the CMZ, suggesting that a central and continuous source of protons with energies up to 1 PeV diffuses out and collides with protons in the ambient gas, producing neutral pi-mesons ($\pi^0$) that decay into observable gamma rays \citep{HESS.2016}. The Fermi bubbles (FBs), two ${\sim}10\,$kpc-scale structures observed to be emanating from the GC by the Fermi Large Area Telescope (Fermi LAT) at  0.1--100\,GeV energies \citep{Su.2010,Ackermann.2014},  are additional evidence of the particle acceleration. To explain them, either inverse Compton emission produced by high energy electrons or the proton--proton interaction just mentioned can be invoked,  both corresponding to particle acceleration from the GC \citep{Ackermann.2014}.

Currently, the Galactic sources of particles with $\approx 1$ PeV energies, so-called PeVatrons, are still under debate \citep{Aharonian.2019}. Supernovae have been proposed as the most important Galactic PeVatron, where CRs are impulsively accelerated on the shock front created by its supersonic outflow \citep{Berezhko.1996,Hinton.2009}. Additionally, stellar winds of Wolf-Rayet (WR) stars in compact and massive clusters of young stars could persistently accelerate CRs due to the collective effect of the shock fronts created by the entire WR population \citep{Klepach.2000}. Finally, accretion onto a compact object could produce high energy particles  \citep{Hinton.2009}. For the CMZ gamma-ray emission, CRs could be accelerating in the vicinity of the supermassive black hole Sgr A* \citep{HESS.2016}, within the Arches (AC), Quintuplet (QC), and nuclear (NSC) clusters of young massive stars \citep{Aharonian.2019},  or due to an unresolved population of millisecond pulsars \citep{Guepin.2018}. Despite many possibilities, the CR origin in the GC is still unknown. Our preliminary results show that the source is inside the NSC if the CMZ is a continuous structure, or from both the NSC and the supernova Sgr A East if the CMZ has an inner cavity \citep[][hereafter Paper~I]{Scherer.2022}.

Additionally, CR transport within the CMZ is not completely understood either. CR transport in the Galactic disk is known to be dominated by diffusion, where the diffusion coefficient is constrained by CR abundance and their fragmentation timescale \citep{Longair.2011,Gaisser.2016}. However, in highly dense zones like the CMZ, it is theoretically expected \citep{Ormes.1988} and constrained from gamma-ray observations \citep{Abeysekara.2017,Aharonian.2019} that diffusion should be two orders of magnitude lower than in the Galactic disk. The main reason is the extreme high-turbulent magnetic fields associated to the high density regions, which create a slow diffusion. Also, the outﬂow velocities of the FBs \citep{Bordoloi.2017} suggest that advective transport towards the Galactic poles may be important for the propagation of CRs within the CMZ. 

The main features of the $\sim$1\,PeV CRs accelerated within the CMZ have been constrained indirectly via gamma-ray flux measurements and considering the gas column density \citep{HESS.2016,HESS.2018}. The gamma-ray luminosity emitted by proton-proton interaction is directly proportional to the ambient gas density \citep{Aharonian.2004, Longair.2011}, but the CMZ 3D morphology has not been completely determined yet \citep{Sofue.2022,Henshaw.2022}. From molecular rotational line emission, e.g. CO lines \citep{Oka.2012} and CS lines \citep{Tsuboi.1999}, different maps of the CMZ column density have been obtained, which have been used as input for hydrodynamical simulations, to quantitatively compare the emission and absorption lines, or to search for the best orbits to explain the observed radial velocities and 2D morphology. From those analyses, a variety of CMZ shapes have been inferred by different studies, including two spiral arms \citep{Sofue.1995,Ridley.2017}, a twisted elliptical ring \citep{Molinari.2011}, an open elliptical stream \citep{Kruijssen.2015}, either an elliptical ring if the star formation is minimal or a fragmented ring if the star formation is intense \citep{Armillotta.2019,Armillotta.2020}, or a bar-like structure \citep{Sawada.2004,Yan.2017}. 

In Paper~I, we explored the impact on the CR analysis of the CMZ having or not a large inner cavity with respect to the line of sight distribution. Using a simple CR model consistent with the \citet{HESS.2016} results, we found that a disk-like gas distribution reproduces the gamma-ray observations considering only stellar winds from WR stars in the NSC. However, a structure with a significant inner cavity requires a composite CR population, coming from the same stellar winds and the supernova Sgr A East. In this paper, we aim to reproduce the HESS observation of the CMZ considering more realistic CR dynamics and CR sources with injection rates derived from their observed kinetic energies. We compute the CR transport considering a differentiated diffusion (inside/outside the CMZ) and polar advection, for both gas distributions proposed in Paper~I, and we explore the CR contribution from the NSC, AC, QC and Sgr A East. Additionally, we show that upcoming observations with the Cherenkov Telescope Array \citep[CTA;][]{CTA.2019} will verify the proposed transport model and constrain its ingredients.

\section{Methodology} \label{sec_metodo} 

We compute a CR transport model using a Monte Carlo method.  We consider a mix of continuous CR sources, namely the NSC, AC and QC, plus impulsive CR injection from Sgr A East. We model transport across two diffusion zones --  low diffusion within the CMZ, and regular Galactic-disk diffusion outside the CMZ. Additionally, we add transport by polar advection. We compute the CR dynamics and calculate the gamma-ray luminosity in a 3D grid, using the same two gas distributions considered in Paper~I -- a CMZ with and without an inner cavity. Finally, we estimate the average CR energy indirectly to find the model that most closely matches the HESS measurements, and propose observations with CTA to confirm and fine-tune the models.

\subsection{Cosmic-ray model}

We model the CR dynamics using the standard diffusion equation for protons \citep{Aharonian.2004, Strong.2007}, neglecting reacceleration as its energy gain is much slower than regular acceleration in shock fronts \citep{Longair.2011,Blasi.2013},
\begin{equation}
\label{eq_dif}
\frac{\partial \psi(\vec{r},p,t)}{\partial t}=\vec{\nabla} \cdot (D(E_p) \cdot \vec{\nabla} \psi)-\vec{\nabla} \cdot (\vec{V} \psi)-\frac{\partial}{\partial p}\left(\left(\frac{ \mathrm{d}p}{\mathrm{d}t}\right)\psi\right)+Q(\vec{r},p,t),
\end{equation} 
where $\psi(\vec{r},p,t)$ is the CR distribution, $\vec{r}$ is the position vector, $p$ is the total CR momentum at position $\vec{r}$, $t$ is the time, $D(E_p)$ is the diffusion coefficient, $E_p$ is the energy of ultrarelativistic particles, $\vec{V}$ is the advection velocity, and $Q(\vec{r},p,t)$ is the CR source term.
In our model, we consider only particles of $E_p=1\,$PeV, so we transfer all energy of CRs above 10 TeV to these mono-energetic particles. We simplify the CR intrinsic spectrum because the solution of Eq. \ref{eq_dif} considered by HESS was obtained with mono-energetic particles subject to standard diffusion. In this way, we can easily compare our new results to those by HESS and also to our own from Paper~I.  When assessing future CTA observations, in Sec.~\ref{sub_CTA}, we distribute the energy across the relevant domain.

\subsubsection{Sources} \label{subsub_source}

We consider three continuous and one impulsive source, with CR injection rates derived from their observed kinetic energies. Additionally, we include another impulsive source whose properties are not  well-constrained observationally. 

\begin{figure*}
\centering
\includegraphics[width=17cm]{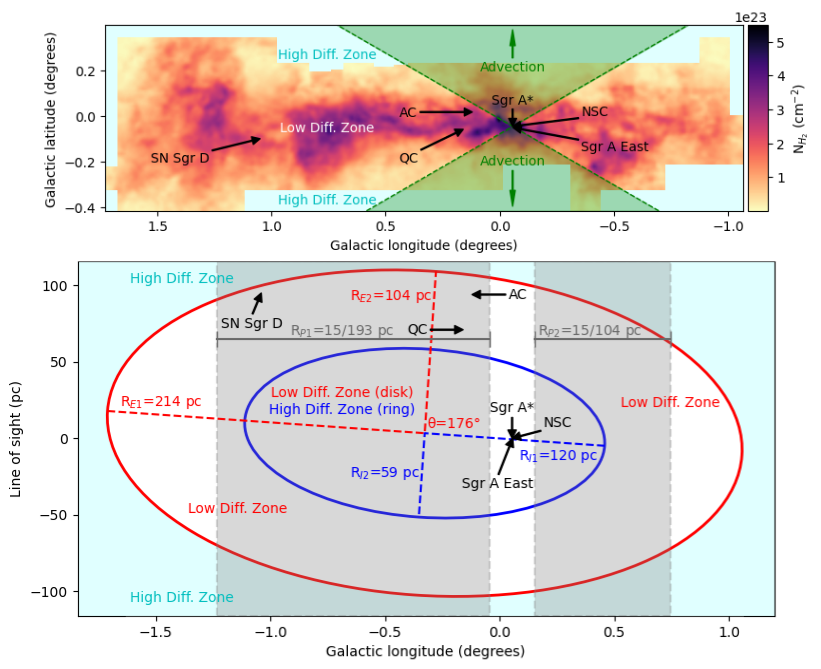}
\caption{CMZ gas distribution in our model, schematic representation of CR dynamics, and CR source location. Top panel: CMZ particle column density spatial distribution computed from CO ($J=3-2$). The CMZ was covered between $-1.07\degr<l<1.73\degr$ and $-0.42\degr<b<0.40\degr$, within a velocity range $|v|< 220$ km s$^{-1}$ \citep{Oka.2012}. The color scale represents the column number density of H$_2$, and a low diffusion of CRs is considered within this region. The cyan outer area corresponds to regions where no gas column density was measured, for which a high diffusion of CRs is computed. The green area denotes the zone with polar advection. Bottom panel: CMZ line of sight distribution boundaries for both the disk and ring configurations, according to Paper~I. The red ellipse is the external boundary for both scenarios, and the blue ellipse is the internal boundary for the ring scenario. A low diffusion of CRs is defined inside the red ellipse for the disk and between the red and blue ellipses for the ring. High diffusion of CRs is computed over the cyan outer area and inside the red ellipse for the ring scenario. R$_{E1}$ and R$_{E2}$ are the semi-major and semi-minor axis for the external ellipse, and R$_{I1}$ and R$_{I2}$ are the semi-major and semi-minor axis for the internal ellipse. Gray zones demarcate the area projected along the line of sight analyzed in the HESS work, where R$_\mathrm{P1}$ and R$_\mathrm{P2}$ are the limits for the projected radii from Sgr A*.}
\label{fig_CMZ_gas}
\end{figure*}

For the continuous sources, which correspond to stellar clusters with colliding WR winds, we define the injection rate as $Q_{con}=A_0(E_p/E_0)^{-\Gamma}g(t)$, where $A_0$ is the CR injection rate, $E_0$ is the energy normalization, $\Gamma$ is the spectral index of the CR intrinsic spectrum, and $g(t)$ is a Heaviside function in the time interval $t_{age}<t$, where $t_{age}$ is the source age. $A_0$ is chosen by assuming that a canonical 1\% of the stellar winds’ kinetic power goes into acceleration of CRs above 10 TeV \citep{Aharonian.2019} and $(E_p/E_0)^{-\Gamma}=1$ for mono-energetic particles with all CR energy above 10 TeV. The proposed continuous CR sources are the following,
\begin{itemize}
    \item NSC: injection rate of $A_0=3.4 \times 10^{36}$ erg s$^{-1}$, considering that the total mass-loss rate of its WR population is $\approx10^{-3}$ M$_\sun$ yr$^{-1}$ and their typical stellar wind velocity is $\approx 1000$ km s$^{-1}$ \citep[see Table 1 of][based on \citealt{Martins.2007}]{Cuadra.2008},
    \item AC: injection rate of $A_0=2.0 \times 10^{36}$ erg s$^{-1}$, considering that the total mass-loss rate of its WR population is $\approx 2.0 \times 10^{-4}$ M$_\sun$ yr$^{-1}$ and their typical stellar wind velocity is $\approx 1850$ km s$^{-1}$ \citep[see Table 2 of][]{Martins.2008},
    \item QC: injection rate of $A_0=1.1 \times 10^{36}$ erg s$^{-1}$, considering that the total mass-loss rate of its WR population is $\approx 8.1 \times 10^{-4}$ M$_\sun$ yr$^{-1}$ and their typical stellar wind velocity is $\approx 670$ km s$^{-1}$ \citep[see Table 2 of][]{Rockefeller.2005b}.
\end{itemize}
For each star cluster, we consider a $t_{age} = 10^6$ yr, since the WR winds remain approximately constant over that timescale \citep{Calderon.2020}, the actual age of the young stellar populations in each cluster is of that order of magnitude \citep{Genzel.2010}, and an accurate assumption of $t_{age}$ makes no difference in this timescale within the CMZ. The source locations are shown in Fig.~\ref{fig_CMZ_gas}, where the line of sight positions of AC and QC are taken from \citet{Hosek.2022}.

For impulsive sources, which correspond to supernova explosions, we define the injection rate as $Q_{imp}=B_0(E_p/E_0)^{-\Gamma} \delta(t-t_{age})$, where $B_0$ is the injection energy and $\delta(t)$ is a Dirac delta function (with units of s$^{-1}$). Again, we assume that a canonical 1\% of the supernova explosion energy goes into acceleration of CRs above 10 TeV \citep{Aharonian.2019} and $(E_p/E_0)^{-\Gamma}=1$ for mono-energetic particles with all CR energy above 10 TeV. The proposed impulsive CR source with known kinetic data is:
\begin{itemize}
    \item Sgr A East: injection energy of $B_0=1.5\times 10^{49}$ erg, considering an explosion energy of $\sim 1.5 \times 10^{51}$ erg \citep{Rockefeller.2005,Fryer.2006}.
\end{itemize}
For Sgr A East, $t_{age} = 1700$ yr \citep{Rockefeller.2005,Fryer.2006}, and its location is shown in Fig.~\ref{fig_CMZ_gas}, where the line of sight coordinate is from \citet{Rockefeller.2005}. 

Finally, motivated by the mismatch between our models and the observations at large radii that we obtained in Paper~I (see also Sect.~\ref{sub_SN_SgrD} below), we consider an additional not well-constrained impulsive source associated with Sgr D. Radio observations have identified three structures close to Sgr D (G1.1–0.1, G1.0–0.2 and G0.9+0.1), which are consistent with supernovae \citep{Heywood.2022}. Some studies indicate that these structures are independent of the CMZ, appearing near it only in projection \citep{Kauffmann.2017}. However, other studies indicate that according to their distances along the line of sight, these structures could be GC objects \citep{Sofue.1995}. Specifically, \citet{Sofue.1995} located G1.1–0.1, G1.0–0.2 and G0.9+0.1 at 8.9 kpc, 8.0 kpc and 9.6 kpc, respectively, with an error of $\approx 30\%$. Given that uncertainty, all of them could be within or outside the CMZ.  We therefore arbitrarily assume that only the middle structure,  G1.1–0.1, is inside the CMZ and located on its far side, as suggested by \citet{Sofue.2022}. So, we model G1.1–0.1 as
\begin{itemize}
    \item SN Sgr D: assumed injection energy of $B_0=5\times 10^{48}$ erg, considering an explosion energy of $5 \times 10^{50}$ erg, which is consistent with the less energetic Galactic supernovae \citep{Acero.2016}.
\end{itemize}
Moreover, for SN Sgr D we assign $t_{age}=10^4$ yr, according to its diameter \citep{Sofue.2022} and indicate its location in Fig.~\ref{fig_CMZ_gas}.

\subsubsection{Differential diffusion}

As the CMZ is a very high density region, a low diffusion coefficient is appropriate. On the other hand, outside of the CMZ conditions should be similar to those in the broader Galactic disc.  Therefore, we consider a low isotropic CR diffusion inside the CMZ defined as $D_{in}(E_p)\sim 10^{26}(E_p/10^{10}$ eV$)^\delta$ cm$^2$ s$^{-1}$ \citep{Ormes.1988,Abeysekara.2017,Aharonian.2019}, and a high isotropic CR diffusion outside the CMZ described by \hbox{$D_{out}(E_p)\sim 10^{28}(E_p/10^{10}$ eV$)^\delta$} cm$^2$ s$^{-1}$ \citep{Aharonian.2004,Strong.2007}. For both regimes, we consider $\delta=0.5$ for a Kraichnan spectrum for magnetic turbulence \citep{Aharonian.2004}.

We define the CMZ borders in Galactic longitude and latitude by the limits of the CO (J=3-2) observation detailed in \citet{Oka.2012}, and the line of sight edges by the distributions proposed in Paper~I, which are shown in Fig.~\ref{fig_CMZ_gas}. The upper panel shows the area without gas observation in cyan, which is considered to have the diffusion coefficient of the Galactic disk, while where the observed column density is shown we use the diffusion coefficient for high densities. The lower panel shows the boundaries of the CMZ as a ring or disk according to Paper~I, where the red ellipse is the outer border of the ring and disk, and the blue ellipse is the inner border of the ring. Within these borders we consider a low diffusion, and outside a high diffusion. 

\subsubsection{Polar advection}

The observation of the FBs suggest polar advective transport. According to \citet{Bordoloi.2017}, the outflow velocity of the FBs is $\approx 1000-1300$ km s$^{-1}$, their age is $\sim 6-9$ Myr, and, at small Galactic latitudes ($b<15\degr$), the FBs have a beam angle of $\approx 120\degr$ from Sgr A*. To quantify the relative importance of advection and diffusion transport it is useful to define a dimensionless number $q = V R_{trans} / D$, where $R_{trans}$ is the transport radius. The CR dynamics are dominated by advection if $q \gg 1$ and dominated by diffusion if $q \ll 1$ \citep{Owens.1977}. In our case $q \approx 0.8$, considering a CMZ thickness of $\approx 75$ pc [$\approx 5\degr$ \citep{Oka.2012} for a GC distance of 8.5 kpc] and the inner diffusion coefficient for 1 PeV particles. As $q$ is somewhat smaller than unity, advection can contribute considerably but not completely dominate the CR dynamics in the area of the FBs.   

In our model, we define the polar advection as a ballistic motion with a velocity of 1000 km s$^{-1}$ towards the north Galactic pole for all particles at an angle of $120\degr$ above Sgr A*, and to the south Galactic pole for all particles at an angle of $120\degr$ below Sgr A*, as is shown in green on the top panel of Fig.~\ref{fig_CMZ_gas}.

\subsubsection{Energy loss} \label{subsub_ener_loss}

$\pi^0$ decay is by many orders of magnitude the dominant process of energy loss for CRs protons \citep{Aharonian.2009, Longair.2011}. This process occurs when high energy protons, or heavier nuclei, collide with protons in the ambient gas. These inelastic collisions produce secondary particles, like $\pi^0$ (e.g., $p+p \rightarrow p+p+\pi^0$), which then decay into gamma-rays \citep[$\pi^0 \rightarrow 2\gamma$;][]{Aharonian.2004,Longair.2011,Gaisser.2016}. However, the particle's average time within the high density volume, i.e. the CMZ, is much shorter than the characteristic cooling time. As such, the proton cooling is neglected in the transport calculation (third term on the right of Eq. \ref{eq_dif}), and only used to generate gamma-ray synthetic maps (Sect.~\ref{sub-gamma-maps}).

\subsubsection{Numerical simulation} \label{subsub-num-sim}

We simulated mono-energetic particles of 1 PeV using the Monte Carlo method in a 3D domain to obtain the modeled CR energy density ($w_{CR}$), where $w_{CR} \propto \psi$. We executed 10 CR simulations of 10$^6$ test particles, for each source and CMZ shape (disk or ring). We discretised the results on a 3D grid centered on Sgr A*, where $w_{CR}$ was computed in each bin. The grid covers Galactic longitudes between $-1.07\degr<l<1.73\degr$, Galactic latitudes within $-0.42\degr<b<0.40\degr$, and $\pm 132$ pc along the line of sight. We assumed a GC distance of 8.5 kpc, with cubic bins of size 4 pc $\times$ 4 pc $\times$ 4 pc, resulting in an angular resolution of $\approx$ 0.03$\degr$. Finally, for each source and CMZ model, all 10 simulations are averaged for every bin.

\subsection{3-dimensional CMZ distribution}

We consider the same 3-dimensional gas distributions as we did in  Paper~I. For more details, see Sect.~2.2 of that paper, which we summarise here.  

We developed two models to distribute the CMZ gas along the line of sight based on \citet{Kruijssen.2015} and \citet{Launhardt.2002}. We consider that the CMZ could be an elliptical ring or an elliptical disk to simulate a CMZ with \citep{Sofue.1995,Kruijssen.2015} and without \citep{Sawada.2004,Yan.2017} an inner cavity. Notice that the latter option includes also models that do consider a cavity \citep{Ridley.2017,Armillotta.2019}, but which is small or shallow enough so it can be neglected for our purposes. Within the disk or ring, we distribute uniformly the column density of molecular gas along the line of sight domain, to obtain an approximation of the 3D CMZ distribution. The total CMZ column density was measured from CO ($J=3-2$) lines observed by the Atacama Submillimeter Telescope Experiment \citep[ASTE;][]{Kohno.2004} and published by \citet{Oka.2012}.\footnote{available at \url{https://www.nro.nao.ac.jp/~nro45mrt/html/results/data.html}} We adopt a mass conversion factor CO-to-H$_2$ of X$_{CO} = 1.8 \times 10^{20}$ cm$^{-2}$ K$^{-1}$ Km$^{-1}$ s \citep{Bolatto.2013} and a ratio of CO($J=3-2$)/CO($J=1-0$) = 0.7 \citep{Oka.2012} at the GC.

The top panel of Fig.~\ref{fig_CMZ_gas} shows the particle column density of the CMZ, while the bottom panel shows its adopted distribution along the line of sight, where the red ellipse is the common external border, and the blue ellipse is the internal border for the ring model. Finally, we computed the gas particle density ($n_\mathrm{H}$) using the same 3D grid described in Sect.~\ref{subsub-num-sim}, considering the gas to be static.

\subsection{Gamma-ray synthetic maps}
\label{sub-gamma-maps}

As described in Sect.~\ref{subsub_ener_loss}, gamma-rays are produced from proton--proton interactions. We compute the gamma-ray luminosity per unit volume ($L_\gamma$/$V_\mathrm{bin}$) in the same grid of Sect.~\ref{subsub-num-sim}, using the values of $w_{CR}$ and $n_\mathrm{H}$ obtained in each grid bin as \citep{Fatuzzo.2006,HESS.2016}
\begin{equation}
\label{eq_lum}
\frac{L_\gamma}{V_\mathrm{bin}} \approx \kappa_\mathrm{\pi}~\sigma_\mathrm{p-p}~c~\eta_\mathrm{N}~n_\mathrm{H}~w_\mathrm{bin},
\end{equation}
where $\kappa_\mathrm{\pi}$ is the fraction of kinetic energy of high-energy protons transferred to $\pi^0$ production, $\sigma_\mathrm{p-p}$ is the cross section for proton-proton interaction and $\eta_\mathrm{N}$ is the gamma-ray contribution from heavier nuclei in CRs and ambient gas. For protons with energies in the  GeV--TeV range, $\kappa_\mathrm{\pi} \approx 0.18$ \citep{Fatuzzo.2006}, for mono-energetic particles of 1 PeV, $\sigma_\mathrm{p-p} \approx 53$ mb \citep{Aharonian.2004}, and $\eta_\mathrm{N} \approx 1.5$ \citep{HESS.2016}. Next, we integrate $L_\gamma$/$V_\mathrm{bin}$ along each line of sight to obtain gamma-ray synthetic maps, where the gamma-ray luminosity per bin cross section is converted a gamma-ray surface flux computed at the Earth reference frame.

Our model neglects the gamma-ray foreground and background of the CMZ \citep{Abramowski.2014} emitted by the isotropic Galactic CRs sea \citep{Blasi.2013}. However, this large-scale emission is considered to be relatively weak and should have a negligible contribution in the central region \citep{HESS.2018}, such that it will not change our main results.

\section{Results}

First, we analyze the results  considering only the sources whose identity and energy is well constrained (i.e. NSC, Sgr A East, AC, and QC). For those sources, we contrast the results with both HESS observations and our Paper~I models. Also, we study the future constraints to be obtained by CTA observations. Finally, we add the proposed injection from SN Sgr D to improve the model at larger Galactic longitude.

\subsection{Model results and current observation}

\begin{figure*}
\centering
\includegraphics[width=17cm]{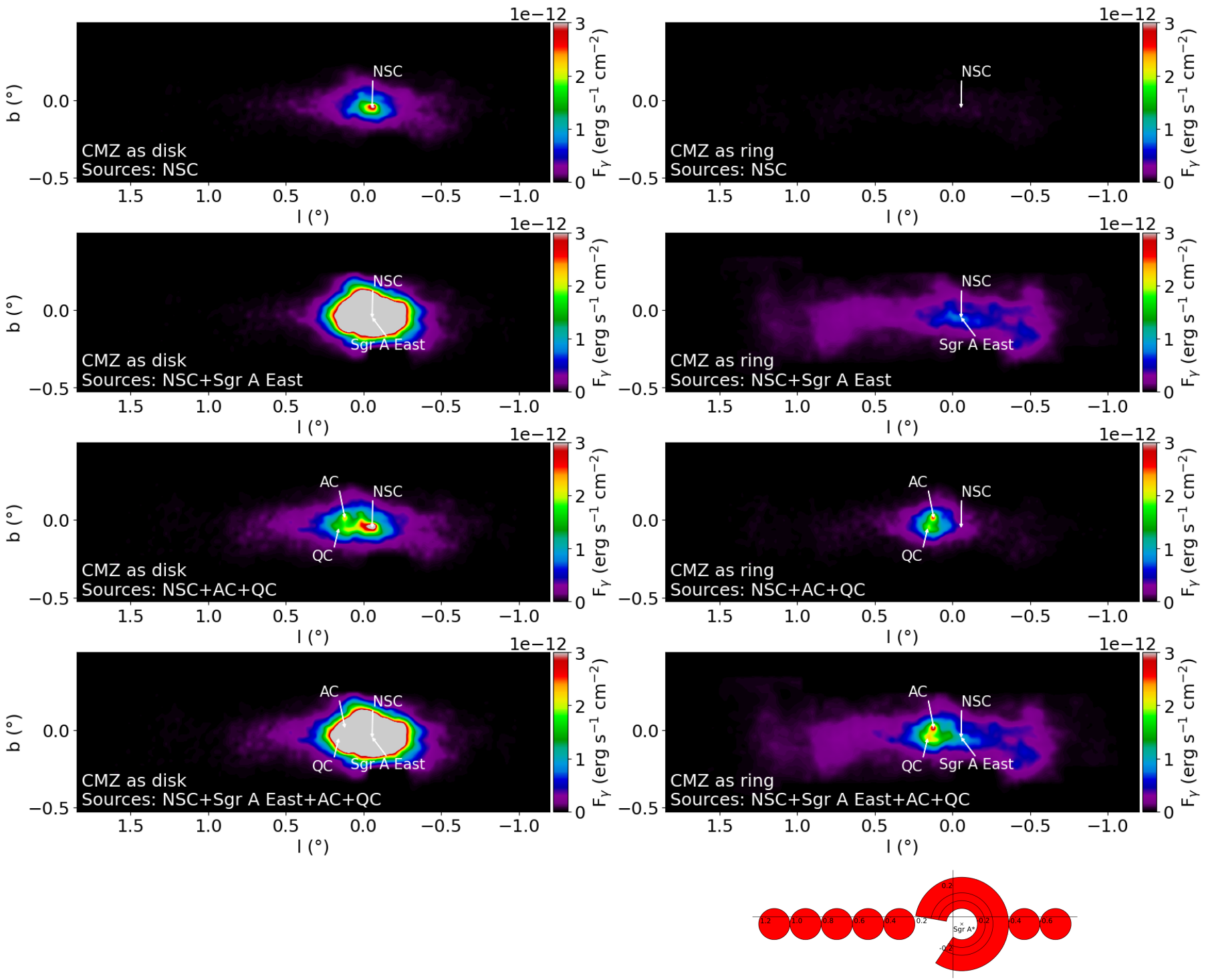}
\caption{Gamma-ray synthetic maps computed as observed from the Earth. Left panels: synthetic maps for the CMZ as a disk and different CR sources (indicated on each panel). Top four right panels:   synthetic maps for the CMZ as a ring and different CR sources (indicated on each panel). All maps have been smoothed with a 0.03$\degr$ Gaussian function to adopt the best CTA beamwidth. Bottom right panel: Regions where we contrast our models with the \citet{HESS.2016} results.}
\label{fig_map_gamma}
\end{figure*}

Figure \ref{fig_map_gamma} shows the results of our models considering the four sources with CR injection derived from observations. Left panels show the results of the CMZ-as-a-disk model, considering NSC (top panel), NSC and Sgr A East (second panel), NSC, AC and QC (third panel), and NSC, Sgr A East, AC and QC (fourth panel) as the CR sources. The top four panels on the right show the result with the same sources but considering the CMZ as a ring. All maps have been adjusted to CTA resolution, i.e. have been smoothed with a 0.03$\degr$ Gaussian function to adopt the best CTA beamwidth.

Comparing all maps, they all show a central overdensity of gamma-ray flux and lower emission at high longitudes. However, models where the CMZ is considered as a disk and Sgr A East is included as a CR source show more emission at larger longitudes. In the case of continuous sources (NS, AC and QC), the particles injected in the past can escape the CMZ due to the external high diffusion, and the gamma-ray emission close to the sources is high because "young" particles have only traveled in the low-diffusion region. In the case of the impulsive source, two different effects are experimented by the particles on either gas configuration. Given the low $t_{age}$ for Sgr A East, the particles are concentrated around the source due to the low diffusion in the disk shape. On the other hand, when the CMZ is a ring, the particles move away quickly, and then get confined when entering the CMZ, as there is not enough time for them to leave again the low-diffusion ring, creating an extended over density. 

\begin{figure*}
\centering
\includegraphics[width=17cm]{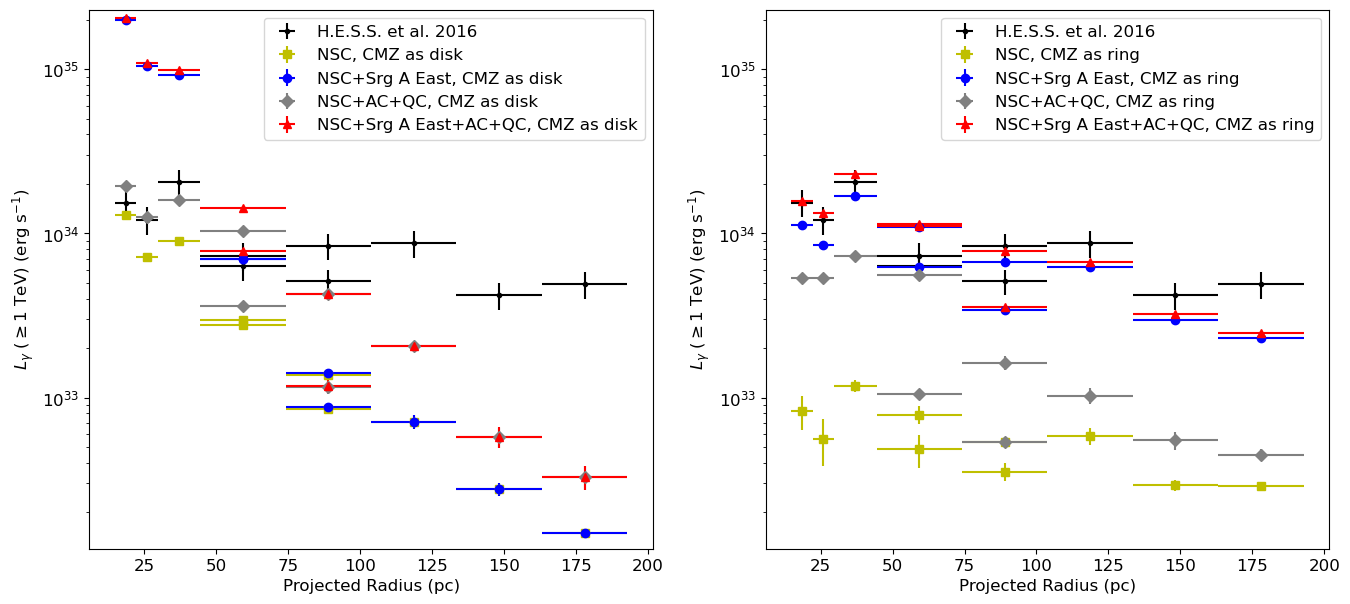}
\caption{CMZ gamma-ray luminosity profiles, computed using the model with realistic CR dynamics introduced in this paper. The luminosities are calculated by integrating the flux in the regions shown in the bottom right panel of Fig. \ref{fig_map_gamma}. Left panel: Gamma-ray profiles considering the CMZ as a disk, for all models shown in the left panels of Fig.~\ref{fig_map_gamma}. Right panel: Gamma-ray profiles considering the CMZ as a ring, for the models shown in the top four right panels of Fig.~\ref{fig_map_gamma}. Black crosses denote the observed gamma-ray luminosity computed by \citet{HESS.2016} in the CMZ. Yellow squares, blue circles, grey diamonds and red triangles, represent the gamma-ray luminosity considering the NSC, NSC+Sgr A East, NSC+AC+QC, and NSC+Sgr A East+AC+QC, as CR sources, respectively.}
\label{fig_lum_all}
\end{figure*}

In order to contrast all models with the HESS observations, we integrate the gamma-ray surface flux over the areas shown in the bottom right panel of Fig.~\ref{fig_map_gamma}. Thus, we obtain gamma-ray luminosities ($L_\gamma$) for the same regions selected in \citet{HESS.2016}. These areas correspond to three annular sectors centred in Sgr A* ($l=-0.056\degr$, $b=-0.04588\degr$), with inner/outer radii of 0.1$\degr$/0.15$\degr$, 0.15$\degr$/0.2$\degr$, 0.2$\degr$/0.3$\degr$, and excluding the region between the angles +10$\degr$ and -56$\degr$ from the positive Galactic longitude axis, plus seven circular regions with 0.1$\degr$ of radius centred in $b=-0.04588\degr$ and $l=-0.656\degr$, -0.456$\degr$, 0.344$\degr$, 0.544$\degr$, 359.344$\degr$, 0.744$\degr$, 0.944$\degr$, and 1.144$\degr$. The comparison of $L_\gamma$ for all models with the HESS data points is show in  Fig.~\ref{fig_lum_all}, where luminosities are radially projected from Sgr A*, therefore two data points are plotted at $\approx 60$ pc and $\approx 90$ pc due to the observation areas selected by HESS. Only the model with the CMZ as a ring and with injection from all sources (NSC, Sgr A east, AC and QC) reproduces satisfactorily the data for projected radii $R \le 150 \textrm{ pc}$. This match is relevant because the CMZ morphology, CR diffusion coefficients, CR polar advection and CR sources have been constrained from observations and previous literature. In conclusion, more realistic CR dynamics suggest that the CMZ presents an inner cavity and that the GC PeVatron can be reproduced by the four modeled sources.

\begin{figure*}
\centering
\includegraphics[width=17cm]{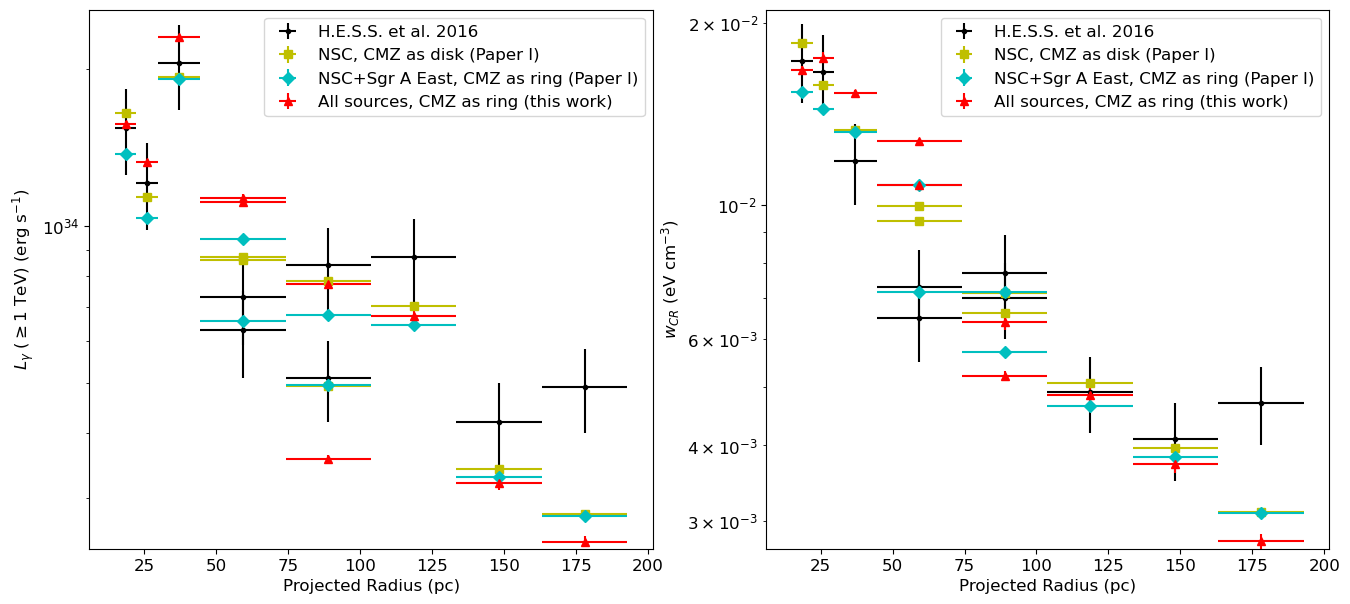}
\caption{Gamma-ray luminosity and average CR energy density for our satisfactory models. Left panel: Gamma-ray luminosity profiles from Sgr A* along the Galactic disk, taken from Fig.~\ref{fig_lum_all} and from Paper~I. Right panel: Corresponding average CR energy densities. Black crosses denote the observed gamma-ray luminosity (left panel) and the indirect observations of CRs (right panel) computed by \citet{HESS.2016}. Yellow squares and cyan diamonds correspond to the CMZ ring and disk models without differential diffusion and polar advection developed in Paper~I. On the disk model, the only CR source is the NSC, while on the ring model the CR sources are the NSC and Sgr A East. Red triangles correspond to the new model with the CMZ as a ring, considering differential diffusion, polar advection, and CR injection from all sources (NSC, Sgr A East, AC and QC).}
\label{fig_comp_lum_CR}
\end{figure*}

In Paper~I, we obtained two scenarios consistent with the \citet{HESS.2016} results but considering a simplified CR transport (i.e. one isotropic diffusion coefficient, no advection, and mono-energetic particles of 1 PeV), which is generally accepted but does not represent properly the GC environment. Here, in the left panel of Fig.~\ref{fig_comp_lum_CR}, we contrast the gamma-ray emission from the best-fitting model of this work with the previous results. Additionally, we compute the average energy density along the line of sight ($w_\mathrm{CR}$) within the same areas used to calculate $L_\gamma$. For that, we consider only the gas column density in the CMZ shown in the top panel of Fig.~\ref{fig_CMZ_gas} (i.e. ignoring the CMZ line of sight distribution, as done typically in the literature), $n_\mathrm{H}=N_\mathrm{H}/l_\mathrm{los}$, and $V=A~l_\mathrm{los}$, where $A$ is the observed gamma-ray area and $l_\mathrm{los}$ is the length of the domain on the line of sight, so Eq. \ref{eq_lum} is rewritten as:  
\begin{equation}
\label{eq_w_cr}
w_\mathrm{CR} \approx \frac{L_\gamma}{\kappa_\mathrm{\pi}~\sigma_\mathrm{p-p}~c~\eta_\mathrm{N}~N_\mathrm{H}~A}.
\end{equation}
In the right panel of Fig.~\ref{fig_comp_lum_CR} we contrast the $w_\mathrm{CR}$ for the CMZ ring and disk models without differential diffusion and polar advection developed in Paper~I, and the model with the CMZ as a ring, considering differential diffusion, polar advection, and CR injection from all sources (NSC, Sgr A East, AC and QC). In both panels, we observe that in general all models are similar and fit the observations,  with only some differences at $R\approx60$ pc, where the new model deviates a bit more. At a projected radius of $R\approx175\,$pc, all models predict a luminosity of a factor of several below the data (see Sect.~\ref{sub_SN_SgrD} below). All in all, considering that we already have these three satisfactory models of Fig.~\ref{fig_comp_lum_CR}, we maintain that fine tuning them is not currently needed. Rather, we now investigate whether future CTA observations will allow us to choose which of the models is the best match.

\subsection{Future observation by CTA} \label{sub_CTA}

CTA will soon be the most powerful telescope for very-high-energy gamma-ray astronomy, covering the energy range from 20 GeV to 300 TeV. CTA will improve the sensitivity by one order of magnitude at 1 TeV compared to current observatories, with an angular resolution of $\approx 0.03\degr$ and an energy resolution of $\approx 8\%$. Very relevant for our study, one of its key science projects will spend 525 hr of exposure on the GC \citep{CTA.2019}. In this context, we now compare the features of the three models that match HESS observations (see Fig.~\ref{fig_comp_lum_CR}) using the expected CTA sensitivity and angular resolution. 

\subsubsection{Energy distribution of synthetic maps}

As explained in Sect.~\ref{sec_metodo}, to calculate the CR dynamics we concentrated all the energy of CRs above 10 TeV into mono-energetic particles of 1 PeV, so our computed gamma-ray luminosity represents the total emission from high energy particles over 10 TeV. To analyze the synthetic maps as real gamma-ray observations, we need to distribute this luminosity in the energy domain, specifically between 1 TeV and 100 TeV, because $E_\gamma \approx 0.1 E_p$ where $E_\gamma$ is the gamma-ray energy \citep{Aharonian.2009,Longair.2011}. \citet{HESS.2016} observed that the CMZ energy spectrum follows a power-law without any cutoff, which was defined as $\mathrm{d}N_{CMZ}/\mathrm{d}E\mathrm{d}A\mathrm{d}t=1.92 \times 10^{-12} (E_\gamma / \mathrm{TeV})^{-2.32}$ TeV$^{-1}$ cm$^{-2}$ s$^{-1}$, where $\mathrm{d}N_{CMZ}/\mathrm{d}E\mathrm{d}A\mathrm{d}t$ is the number of gamma-rays emitted from the CMZ per unit of energy, area and time, as a function of $E_\gamma$. Then, we define the total gamma-ray flux ($F_\gamma$) as \citep{HESS.2018b}
\begin{equation}
\label{eq_flux_bin}
F_\gamma=\Phi_0 \int_{1~\mathrm{TeV}}^{100~\mathrm{TeV}} E\frac{\mathrm{d}N_{CMZ}}{\mathrm{d}E\mathrm{d}A\mathrm{d}t} \,\mathrm{d}E ,
\end{equation}
where $\Phi_0$ is the flux normalisation. Finally, we discretized all $F_\gamma$ of each pixel over five bins per energy decade and adjusted $\Phi_0$ for every $F_\gamma$. 

\subsubsection{Gamma-ray absorption}

Gamma-ray emissions can be absorbed during their propagation in the Galaxy by pair production. Specifically, high-energy photons above $\approx 10$ TeV and $\approx 200$ TeV interact with the Galactic interstellar radiation field (ISRF) and the cosmic microwave background (CMB), respectively, producing an electron-positron pair \citep[$\gamma+\gamma \rightarrow e^- + e^+$;][]{Moskalenko.2006,Popescu.2017}. Therefore, we attenuated $F_\gamma$ according to the gamma-ray transmittance (i.e. the fraction of photons reaching the Earth) as a function of $E_\gamma$ for gamma-ray sources located at the GC computed by \citet{Moskalenko.2006}, where transmittance decreases to $\approx 0.8$ for gamma-rays between 50--100\,TeV. 

\subsubsection{Data analysis of synthetic maps} \label{sub_data_cta}

According to the methodology described in \citet{Mohanty.1998}, we convert $F_\gamma$ to photon counts following
\begin{equation}
\label{eq_flux_count}
N_{ON}=\Delta t \int_{1~\mathrm{TeV}}^{100~\mathrm{TeV}} \Phi_0 
 \frac{\mathrm{d}N_{CMZ}}{\mathrm{d}E\mathrm{d}A\mathrm{d}t} A_0 \epsilon_\gamma  \,\mathrm{d}E ,
\end{equation}
where $N_{ON}$ is the number of gamma-rays observed by CTA, $\Delta t$ is the observation time, $A_0$ is the telescope effective area, and $\epsilon_\gamma$ is the data taking efficiency. For the CTA telescope, $A_0=5 \times 10^{10}$ cm$^2$, $\epsilon_\gamma=0.7$, and we considered the typical observation time to be 50 hr \citep{CTA.2019}. We computed $N_{ON}$ for each modeled pixel of angular resolution $0.027 \degr$. To estimate the background, we use the GC gamma-ray observation of the H.E.S.S. Galactic plane survey \citep{HESS.2018b} and the field-of-view background method, which is appropriate for extended sources \citep{Berge.2007}. We select an area centred on Sgr A* of $4.5\degr$, which is equal to the minimum CTA field of view \citep{CTA.2019}, and excluded known gamma-ray sources under the assumption that they will be easily isolated by CTA. From the entire resulting area, we obtained a background count of $N_{OFF}=259006$ in a total solid angle of $\Omega_{OFF}=0.00273$ sr for the same 50 hr of observation. Finally, we estimate the number of excess events over the background ($N_\gamma$) for each pixel as \citep{HESS.2018b} 
\begin{equation}
\label{eq_N_cout}
N_{\gamma}= N_{ON}- \alpha N_{OFF}, 
\end{equation}
where $\Omega_{ON}$ is the pixel solid angle and $\alpha=\Omega_{ON}/\Omega_{OFF}=8 \times 10^{-5}$. Then we compute the gamma-ray statistical significance for each event ($N_{ON}$, $N_{OFF}$) according to \citet{Li.1983}, where the event standard deviation ($\sigma$) is defined by
\begin{equation}
\label{sigma}
\begin{split}
&\sigma=\sqrt{2 \left[ N_{ON} ~\mathrm{ln}\left( a \right) + N_{OFF} ~\mathrm{ln}\left( b \right) \right]},\\
&a=\left(\frac{1+\alpha}{\alpha}\right) \frac{N_{ON}}{N_{ON}+N_{OFF}},\\
&b=\left(1+\alpha\right) \frac{N_{OFF}}{N_{ON}+N_{OFF}}.\\
\end{split}
\end{equation}

To estimate the influence of the systematic uncertainty and statistical error, we compute the signal-to-background ratio ($SBGR$) and the signal-to-noise ratio ($SNR$) as
\begin{equation}
\label{eq_SBR}
SBGR= \frac{N_{\gamma}}{N_{bg}} ,
\end{equation}
\begin{equation}
\label{eq_SNR}
SNR= \frac{N_{\gamma}}{\sqrt{N_{CR}^2+N_{CTA}^2}},
\end{equation}
where $N_{bg}=\alpha N_{OFF}$, $N_{CR}$ is the uncertainty in counts associated to the  statistical error of the CR model, and $N_{CTA}$ is the uncertainty in counts associated to the telescope systematic error. Finally, we consider that gamma-ray emission is actually detected when $N_{\gamma} \geq 10$, $\sigma \geq 5$, $SBGR \geq 0.05$ and $SNR \geq 5$ \citep{Actis.2011}, considering an uncertainty in the background of 1\% \citep{Funk.2013}. 

\begin{figure*}
\centering
\includegraphics[width=17cm]{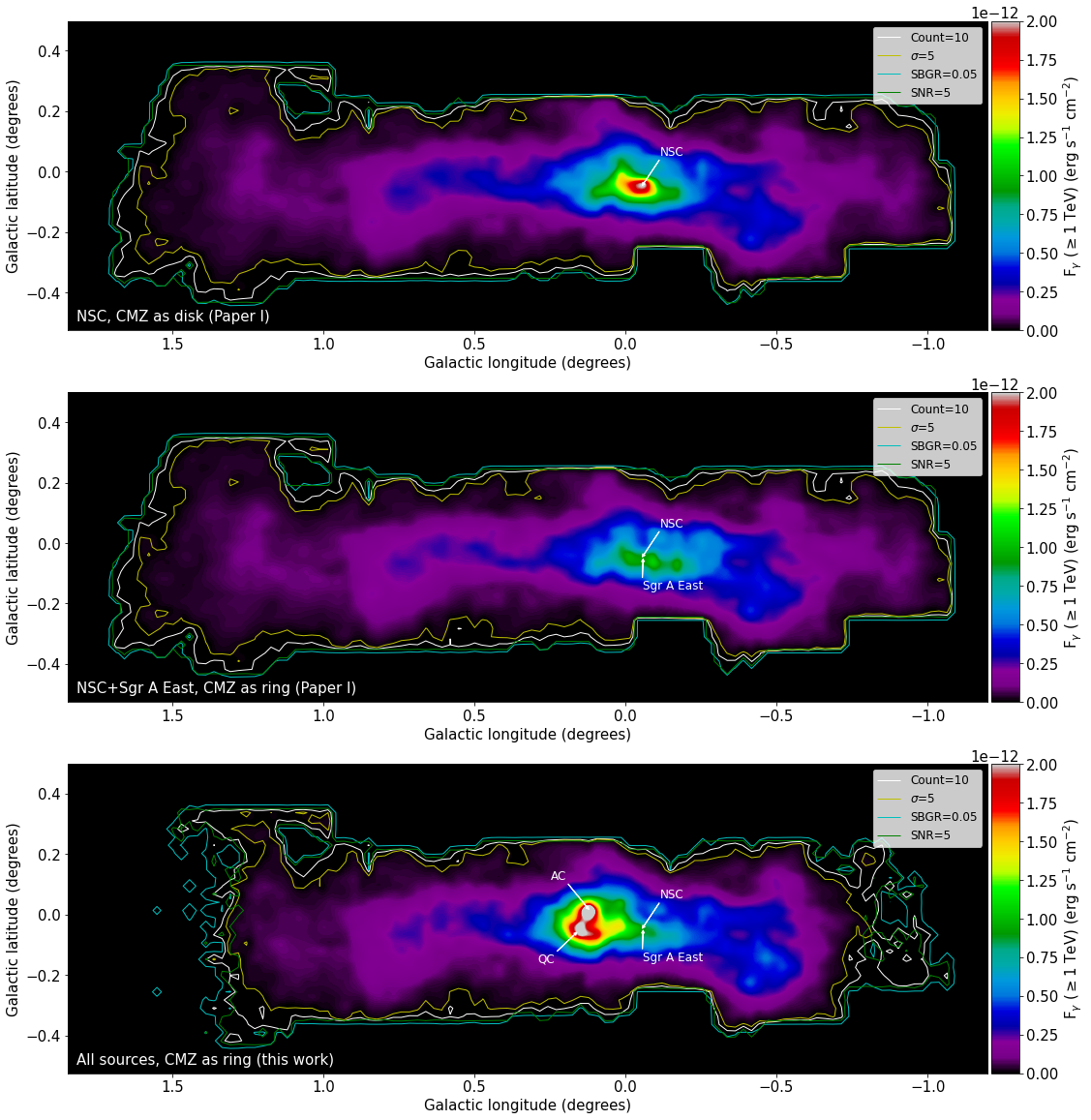}
\caption{Synthetic maps of gamma-ray flux, considering 50 hr of observation with CTA and $E_\gamma \geq 1$ TeV. Top and middle panels: Gamma-ray flux of CMZ models without differential diffusion and polar advection, reported in Paper~I. Bottom panel: Gamma-ray flux of CMZ model with differential diffusion and polar advection. CR sources and the CMZ morphology are indicated on each panel. White, yellow, cyan and green contours are lower limits to $N_{\gamma}$, $\sigma$, $SBGR$ and $SNR$, respectively. All maps have been smoothed with a $0.03\degr$ Gaussian function to adopt the CTA beamwidth.}
\label{fig_Flux_1TeV}
\end{figure*}

\begin{figure*}
\centering
\includegraphics[width=17cm]{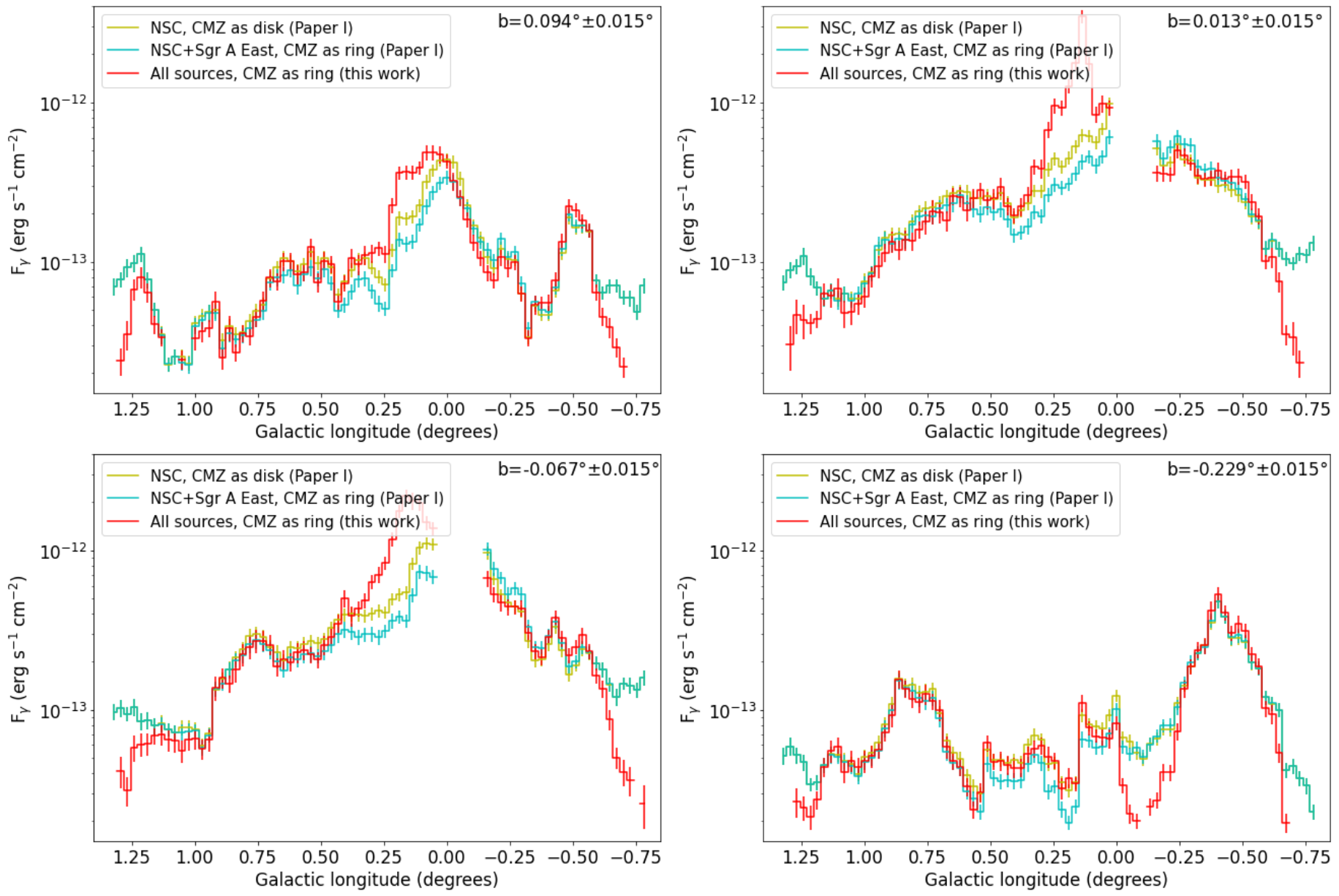}
\caption{Surface gamma-ray flux profiles for successful models. Profiles are obtained between $0.797\degr<l<1.337\degr$ and $b=0.094\degr$ (top-left panel), $0.013\degr$ (top-right panel), $-0.067\degr$ (bottom-left panel) and $-0.229\degr$ (bottom-right panel). The profile is binned adopting the CTA angular resolution, $0.03\degr$. Lines show gamma-ray flux considering 50 hr of observation with CTA and $E\gamma \geq 1$ TeV. Yellow and cyan profiles represent the models proposed in Paper~I with the CMZ as a disk and a ring. The red profiles are from the model considering differential diffusion, polar advection, and CR injection from NSC, Sgr A East, AC, and QC (all sources). The central $0.2\degr$ around Sgr A* was excluded.}
\label{fig_F-pro}
\end{figure*}

Figure \ref{fig_Flux_1TeV} shows the synthetic maps of gamma-ray surface flux, considering 50 hr of observation with CTA and $E_\gamma \geq 1$ TeV, where contours demarcate the detection criteria. The clearest differences between the maps are with regard to the extension and the CR sources. In order to compare in detail our models, Fig.~\ref{fig_F-pro} shows selected longitudinal profiles with prominent features, where $F_\gamma$ was obtained between $-0.797\degr<l<1.337\degr$ and for $b=0.094\degr$, $0.013\degr$, $-0.067\degr$ and $-0.229\degr$. A $\Delta l$ and  $\Delta b$ of $0.03\degr$, comparable to the CTA beamwidth, is adopted for all cases. In addition, we do not consider the central $0.2\degr$ around Sgr A* because HESS observes here a point-like source with a different energy spectrum than the CMZ \citep{HESS.2016}. Moreover, the amount of gas in this region, including the circumnuclear disk \citep{Genzel.1988}, is still under debate \citep{Chen.2016}. In all panels, the differential diffusion effect is observed at $l \approx 1.2\degr$ and $l \approx -0.6\degr$, where $F_\gamma$ decrease due to the external high diffusion. The effect is similar in both directions but at a different relative radius with respect to Sgr A* ($l \approx -0.05\degr$),  due to the asymmetric distribution of the CMZ gas. In the bottom-right panel, the polar advection effect becomes noticeable, where $F_\gamma$ decreases at Sgr A* for the model that includes advection and a roughly flat profile is observed for the other models. This is due to advection allowing CRs to escape faster than in other areas of the CMZ. When the CR dynamics is dominated by diffusion or advection it will decrease $\propto 1/r$ or $\propto 1/r^2$ respectively \citep{HESS.2016}, such that the effect is more prominent farther away from Sgr A* and NSC. In the top-right and bottom-left panels, the CR injection from AC ($l \approx 0.12\degr$ and $b \approx 0.02\degr$) and QC ($l \approx 0.16\degr$ and $b \approx -0.06\degr$) are resolved as $F_\gamma$ peaks. On the other hand, the models without differential diffusion and polar advection show a central small bump at different longitudes ($l \approx -0.05\degr$ for the disk model and $l \approx -0.08\degr$ for the ring), which traces the central $w_\mathrm{CR}$ peak. This discrepancy is due to their respective gas distributions along the line of sight.

\subsection{CMZ gamma-rays over 50 TeV}

\begin{figure*}[ht]
\centering
\includegraphics[width=17cm]{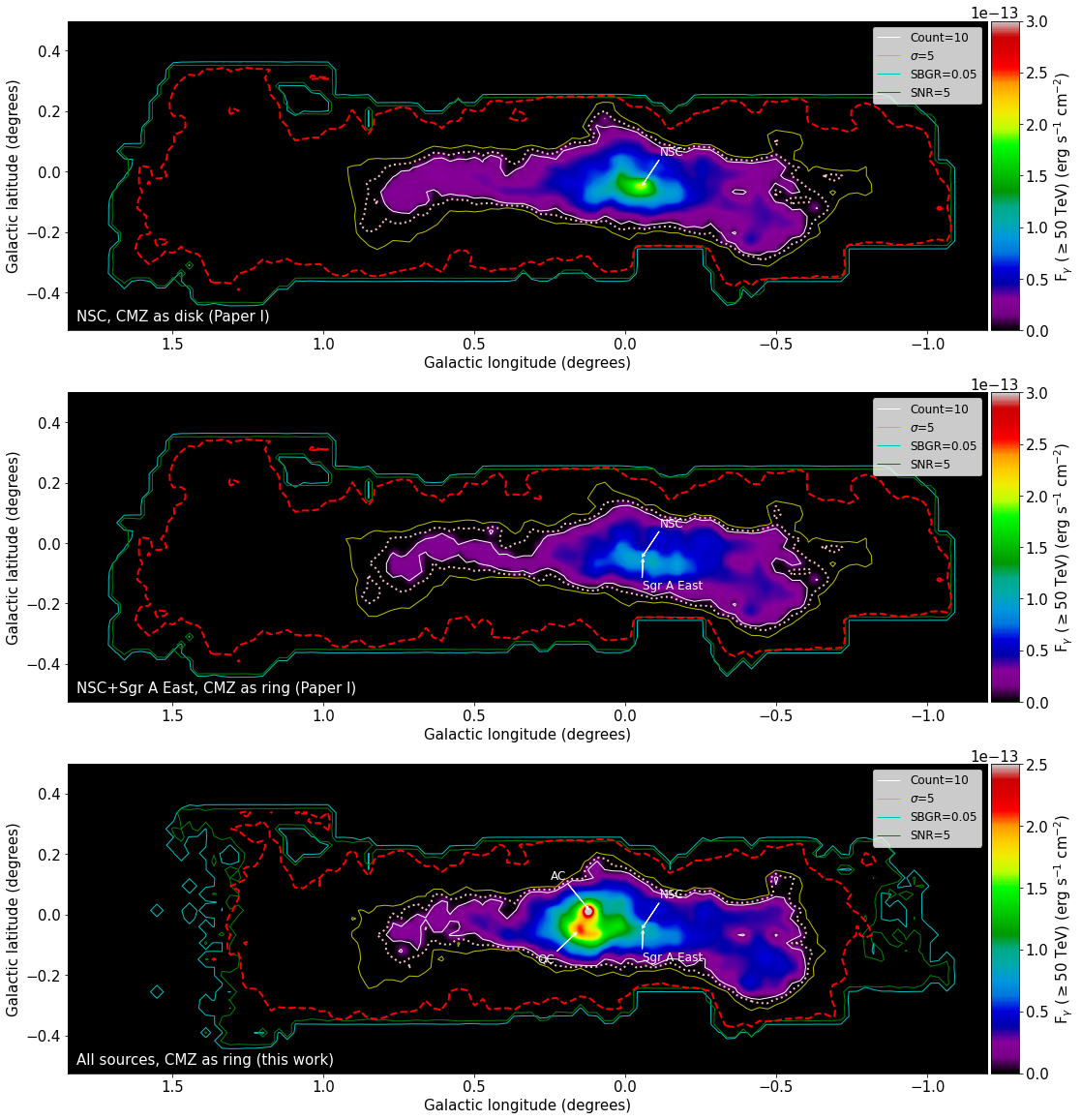}
\caption{Same as Fig.~\ref{fig_Flux_1TeV}, but considering 525 hr of observation with CTA and $E_\gamma \geq 50$ TeV. The additional pink dotted contours correspond to the maps perimeter without considering gamma-ray absorption, while red dashed contours mark the perimeter of the synthetic maps of Fig.~\ref{fig_Flux_1TeV}.}
\label{fig_Flux-50TeV}
\end{figure*}

The CR spectrum at PeV energies is still not well-constrained observationally.  Although the HESS analysis is consistent with protons of $\sim$ PeV, observation of gamma-rays close to 100 TeV would be unequivocal evidence for a GC PeVatron.
Since such particle acceleration is assumed in our model, we analyzed if CTA would detect gamma-rays over 50 TeV. At gamma-ray energies over $\approx 10$ TeV, the photon detection method is dominated by $N_{\gamma} \geq 10$, where the telescope array needs to cover $\geq$ km$^2$ \citep{Funk.2013} or a very long observation time is needed (Eq. \ref{eq_flux_count}). In this context, \citet{HESS.2016} did not observe any gamma-rays over $\approx$ 50 TeV with $\approx$ 200 hr of exposure time, while our models with 50 hr of observation with CTA are dominated by photons between 1 TeV and 10 TeV. 
To address this issue, we repeat the data analysis (see Sect.~\ref{sub_data_cta}), but only considering gamma-rays above 50 TeV, 525 hr of observation with CTA, and a background that follows a power-law with a spectral index of -2.57 \citep{Bernlohr.2013}. Figure \ref{fig_Flux-50TeV} shows the gamma-ray flux as observed by CTA for photons over 50 TeV. In all models, the CMZ is partially filled by gamma-rays and extended emission is observed in the central area, where gamma-rays between 50 -- 100 TeV are detected even considering the gamma-ray absorption by pair production. CTA will be able to trace CRs of $\approx$ PeV energy, and their detection will cover the area of all proposed sources, with AC and QC being individually resolved.  

Therefore, CTA will directly confirm whether clusters of young massive stars constitute PeVatrons. Such a detection would provide relevant information about the Galactic origin and acceleration of CRs up to 1 PeV without all the complications of the NSC. In the latter a point-like gamma-ray source is observed \citep{HESS.2016}, but its origin remains ambiguous due to other potential options: Sgr A* could inject CRs \citep{Aharonian.2005}; gamma-ray emission from a dark matter peak is theoretically expected \citep{Belikov.2012}; an unresolved population of millisecond pulsars could accelerate particles \citep{Guepin.2018, Bower.2018}; there is a pulsar wind nebula on the line of sight \citep{Wang.2006}; stellar winds could be interacting further with an outflow coming from the supermassive black hole vicinity \citep{Cuadra.2015,Russel.2017}; and we suggested an additional CR injection from Sgr A East (Paper~I).  

\subsection{Gamma-rays from Sgr A*} \label{sub_SgrA*}
As mentioned previously, \citet{HESS.2016} reported a point-like source consistent with emission from the location of Sgr A*. The main difference between Sgr A* observation and the gamma-rays from the CMZ is the energy cutoff at 10 TeV observed from this point-like source. We integrate the gamma-ray surface flux over the same area that was observed by HESS (i.e. a circular region of radius 0.1$\degr$ centered in Sgr A*), and our model predicts a luminosity four times lower than the luminosity obtained from integrating the observed spectrum between 1 - 100 TeV. Also, if we consider that our prediction follows the CMZ observed spectrum, the flux magnitude of the cutoff is above our modeled luminosity. 

We explored including a new CR source at the location of Sgr A* so that the model better matches the point-like source observation, however,
any additional CR population modified strongly the profiles shown in Fig \ref{fig_comp_lum_CR} when interacting with the CMZ gas. Therefore, our NSC+Sgr A East+AC+QC model (with the CMZ as a ring) requires an additional gamma-ray source with an energy cutoff at 10 TeV, an extension smaller than 0.2°, and located at the Sgr A* coordinates.

\subsection{CR injection from SN Sgr D} \label{sub_SN_SgrD}

\begin{figure*}[ht]
\centering
\includegraphics[width=17cm]{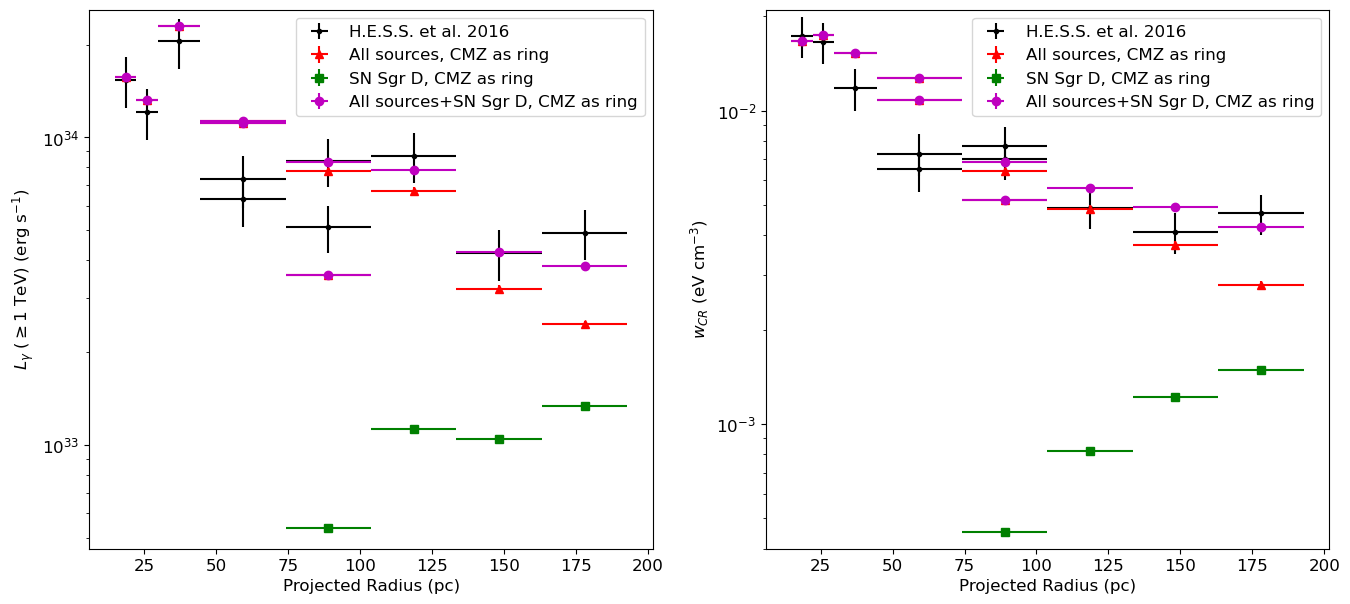}
\caption{Same as Fig.~\ref{fig_comp_lum_CR}, but now plotting the gamma-ray luminosity and the indirect observations of CRs for the CMZ ring, considering differential diffusion, polar advection, and CR injection from SN Sgr D (green squares). Magenta circles are the sum of both components (all sources+SN Sgr D).}
\label{fig_Best+SgrD}
\end{figure*}

In Paper~I, we attributed the discrepancy at $R\approx175\,$pc to more realistic CR dynamics and/or an additional CR source. As is evident from Fig.~\ref{fig_comp_lum_CR}, more realistic dynamics did not solve this problem. Therefore we analyzed the possible injection of CRs from SN Sgr D to fit this detection. In Fig.~\ref{fig_Best+SgrD}, we add to the NSC+Sgr A East+AC+QC model (CMZ as a ring) the SN Sgr D simulation. The sum reproduces the observed gamma-ray emission satisfactorily and improves our model, although assuming arbitrary yet reasonable values for the energy of the explosion and the line-of-sight distance for that source.

 To estimate the impact of adding SN Sgr D to our best model with differential diffusion and polar advection,  we recompute the synthetic maps and profiles observed by CTA, finding that SN Sgr D does not modify our previous analysis. The maps and profiles considering SN Sgr D are shown in Figs.~\ref{fig_Flux-SgrD-appen} and \ref{fig_Lon-D-appen}. 

\section{Discussion}

In Paper~I, we found two models for the CR dynamics in the GC that were consistent with HESS gamma-ray observations. Those models however were too simplistic, and neglected several expected physical effects, such as differential diffusion and polar advection, and CR sources, like the Arches and Quintuplet stellar clusters. These ingredients were included in the models presented here, and we obtained a new model that also recreates the HESS data. This match is remarkable, as it is based on a priori reasonable assumptions, without fine-tuning any parameters. For all physical processes and sources we used fiducial values based on observations and/or the current literature. However, other combinations of sources and processes could also fit the data, and of course there are uncertainties in the fiducial values we used (e.g. the rate of kinetic energy going into acceleration of CRs above 10 TeV). Therefore, we do not claim that our model corresponds to the reality, and instead simply argue that upcoming CTA observations will be able to constrain the models in unprecedented detail.

According to our observational analysis, the different HESS-matching models will show different features when imaged by CTA. The CR realistic transport model proposed in this work will present the most different signature, where CTA will be able to constrain the differential diffusion effect close to the CMZ edge and verify the polar advection to high or low latitudes at the longitude of Sgr A*. Moreover, thanks to its fine angular resolution of $\approx$ 0.03$\degr$ \citep{CTA.2019}, it will be possible to detect the gamma-ray peaks associated to AC, QC, and a composite source of both NSC and Sgr A East, independently, as well as any other potential CR acceleration sites so far not considered. Additionally, using these future observations, we could fine tune our model to determine observationally the CMZ diffusion coefficient, polar advection magnitude, the source injection rates, and the gas distribution along the line of sight.

Regarding gamma-ray emission over 50 TeV, we verified the main results for more strict transmittances \citep[e.g. $\approx 0.75$ for gamma-rays between 50 – 100 TeV;][]{Popescu.2017,Porter.2018}. The conclusions are the same, CTA will directly confirm PeVatrons in the GC, and the observed morphology will only turn out to be less extended due to more intense ISRF. Additionally, the spectrum that CTA will observe should be corrected due to gamma-ray absorption to obtain the intrinsic CR spectrum to PeV energies. 

As shown above, our model remains consistent with the current observations of Sgr A*, and requires an additional point-like gamma-ray source at that location. This emission has been argued to arise from the vicinity of the supermassive black hole \citep{Aharonian.2005} or the ambient gas in the inner few pc \citep{Chernyakova.2011}, within our modeled cavity.

While almost all our sources are already well-constrained from observations, SN Sgr D remains poorly determined. Its currently assumed properties are consistent with its diameter \citep{Sofue.2022} and with the energetics of the weaker supernovae in our galaxy \citep{Acero.2016}. According to our simulations, a verification of this source by CTA may be very difficult due to its low CR injection rate. Additionally, a more accurate determination of its line-of-sight location is necessary.

\begin{figure*}[ht]
\centering
\includegraphics[width=17cm]{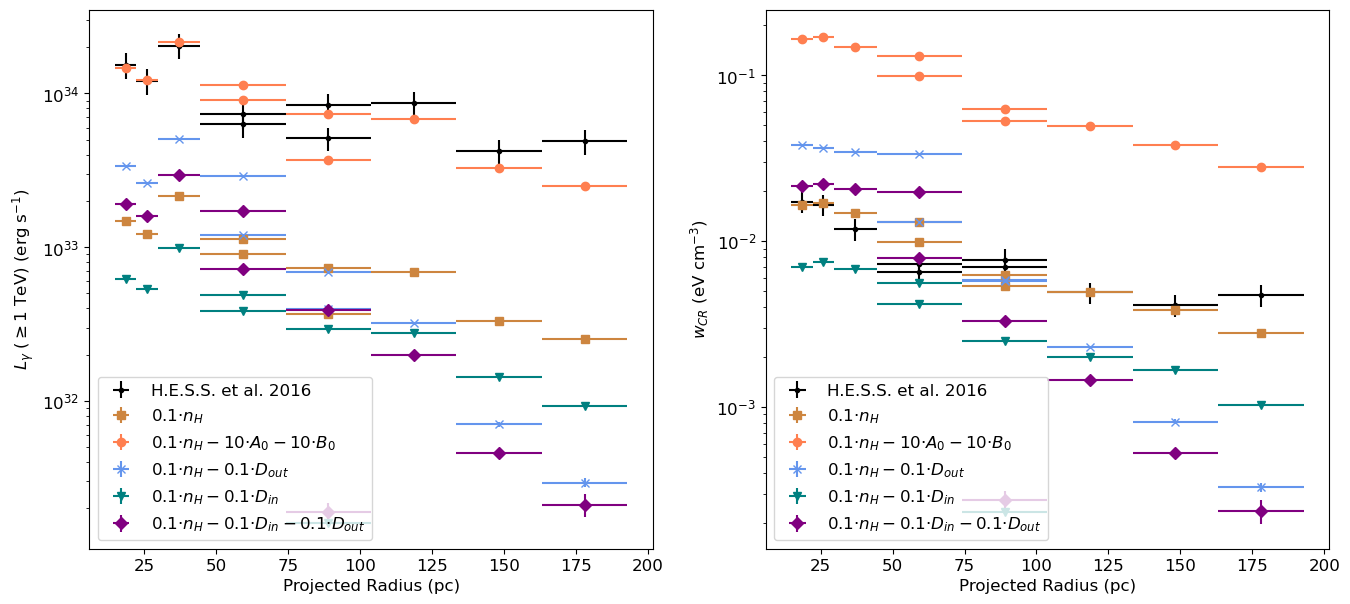}
\caption{Same as the model of all sources and CMZ as a ring in Fig.~\ref{fig_comp_lum_CR}, but setting the CO-to-H$_2$ mass conversion factor to one order of magnitude lower than the Galactic value. We compute models considering the same previous dynamics and CR injection (brown squares), ten times more CR injection (orange circles), ten times less diffusion coefficient outside the CMZ (light blue x-marks), ten times less diffusion coefficient inside the CMZ (green triangles), and ten times less diffusion coefficient inside/outside the CMZ (purple diamonds).}
\label{fig_Mod_sce}
\end{figure*}

The model presents some limitations and uncertainties, which could affect our result. Most importantly, we followed \citet{HESS.2016} and used the standard Galactic CO-to-H$_2$ mass conversion factor for the CMZ. However, other studies estimate it to be lower by up to an order of magnitude \citep{Sodroski.1995,Dahmen.1998, Sofue.2022}. In Fig. \ref{fig_Mod_sce}, we recompute the NSC+Sgr A East+AC+QC model (CMZ as a ring) setting the CO-to-H$_2$ mass conversion factor to X$_{CO} = 1.8 \times 10^{19}$ cm$^{-2}$ K$^{-1}$ Km$^{-1}$ s, i.e., one order of magnitude lower than the Galactic value, considering the same CR dynamics and injection as our satisfactory model from Sect.~\ref{sec_metodo} (brown squares), a ten times higher CR injection rate (orange circles), a ten times lower diffusion outside the CMZ (light blue x-marks), a ten times lower diffusion inside the CMZ (green triangles), and a ten times lower diffusion both inside and outside the CMZ (purple diamonds). Analyzing these models, it is evident that the CR energy density inferred by HESS is in strong tension with the lower value of the gas conversion factor and therefore strong additional sources would be required to match the gamma-ray observations. For instance, the model with the same sources but ten times higher injection rates, produces the same luminosity as in our fiducial model.  We neglected additional potential CR sources, such as the Sgr B1 and Sgr B2 star formation complexes \citep{Ginsburg.2018,Henshaw.2022} or other supernova remnants that are visible in radio and X-ray maps \citep[e.g.,][]{Heywood.2022}, because our aim in this study was to compute a fiducial model supported by known CR sources with well-constrained kinetic energies and low uncertainty on whether they are within the CMZ \citep{Crocker.2007,Protheroe.2008,Acero.2016,Kauffmann.2017}. In future works, we will explore the impact of a more exhaustive characterization of these potential sources.  On the other hand, considering only the known sources, the CMZ mass conversion factor would need to be close to the standard Galactic one. 

Other uncertainties include that the positions of AC and QC along the line of sight are not well constrained. However, repeating the simulations with the locations proposed by \citet{Kruijssen.2015} on the near side of the CMZ, we find that the results do not vary substantially, as the clusters are still embedded in high-density gas. Additionally, CR diffusion and advection areas were delimited from molecular emission lines and gamma-ray observations, respectively, which can only approximate the actual CR dynamics. Moreover, the transition between zones could be smoother.

Finally, our results are consistent with recent theoretical models of the CMZ and Westerlund 1  \citep[a cluster of young massive stars with a gamma-ray emission similar to the CMZ;][]{Aharonian.2019}. For the CMZ, a composite population of CRs within the GC was constrained to reproduce current observations \citep{Becker.2022}, and for Westerlund 1, CR acceleration at the wind termination shock within massive stars clusters and a low diffusion in high density zones were derived to be consistent with gamma-ray observations \citep{Bhadra.2022}.

\section{Conclusion}

We developed a Monte Carlo CR diffusion model for the GC CRs considering differential diffusion, polar advection and several sources. We studied the gamma-rays emitted via proton-proton collision with the CMZ gas, creating synthetic gamma-ray maps. We considered two different CMZ 3D models, with either a ring or a disk shape, that simulate a significant inner cavity or not. Our simulation results are consistent with the former, implying that the CMZ has an inner cavity and that the CR injection is from the combination of the NSC, Sgr A East, AC and QC. The resulting gamma-ray luminosity was found using fiducial parameters and normalizations derived from observations and the current literature, including the standard Galactic CO-to-H$_2$ conversion factor. If a CO-to-H$_2$ conversion factor one order of magnitude lower is considered, the same sources would need ten times higher injection rates to match the observed gamma-ray luminosity. Additionally, an impulsive emission from SN Sgr D can explain the relatively high gamma-ray emission detected by HESS at a large projected radius from Sgr A*. Our conclusion on the shape of the CMZ is consistent with several kinematic and dynamical gas models that require an inner cavity, but stands in contrast to models based on observations of molecular lines and some dynamical gas models, which suggest a continuous structure. Finally, our predictions do not overproduce the current gamma-ray spectrum observed from Sgr A*, for which additional physical mechanisms have been proposed.

To verify whether this model, or the models without differential diffusion and polar advection propounded in Paper~I, explain better the gamma-ray emission, we studied the signatures that CTA could detect in the future. Considering 50 hr of observation, CTA will distinguish those models, constraining the effects of differential diffusion, polar advection, CR sources, and the CMZ morphology. Moreover, with the planned total 525 hr of observation, CTA should detect gamma-rays close to 100 TeV, which will be unequivocal evidence of a GC PeVatron, and even confirm whether clusters of young massive star without a super-massive black hole can accelerate particles up to 1 PeV.

\begin{acknowledgements}
We thank the anonymous referee, and AS's thesis reviewers, Mario Riquelme and Rolando D\"unner, for constructive comments that helped us improve the paper. This project was partially funded by the Max Planck Society through a “Partner Group” grant. AS acknowledges the help and useful comments by Brian Reville, the hospitality of the Max Planck Institute for Nuclear Physics, where part of the work was carried out, and funding from the Deutscher Akademischer Austauschdienst (DAAD). We thank Roberto Lineros for useful discussions at an early stage of this project. AS and JC acknowledge financial support from FONDECYT Regular 1211429. FEB acknowledges support from ANID-Chile BASAL CATA FB210003, FONDECYT Regular 1200495 and 1190818, and Millennium Science Initiative Program – ICN12\_009. The Geryon cluster at the Centro de Astro-Ingenieria UC was extensively used for the calculations performed in this paper. BASAL CATA PFB-06, the Anillo ACT-86, FONDEQUIP AIC-57, and QUIMAL 130008 provided funding for several improvements to the Geryon cluster.
\end{acknowledgements}

\bibliographystyle{aa}
\bibliography{bi}

\onecolumn
\begin{appendix}

\section{Maps and profiles considering SN Sgr D}

\begin{figure*}[h]
\centering
%\sidecaption
\includegraphics[width=15.5cm]{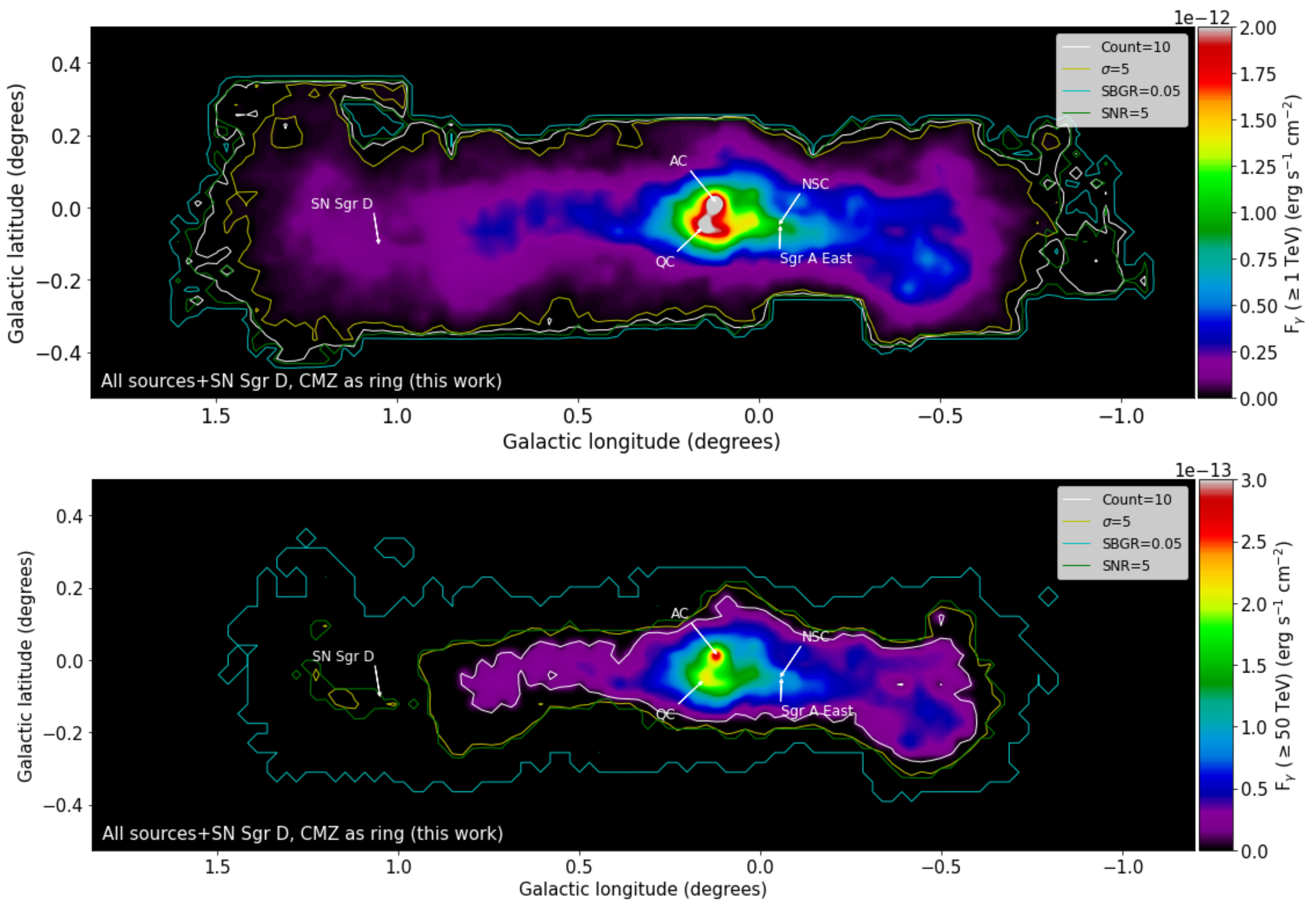}
\caption{Same as bottom panel of Figs. \ref{fig_Flux_1TeV} (left panel) and \ref{fig_Flux-50TeV} (right panel), but now adding SN Sgr D.}
\label{fig_Flux-SgrD-appen}
\end{figure*}

\begin{figure*}[h]
\centering
%\sidecaption
\includegraphics[width=15.5cm]{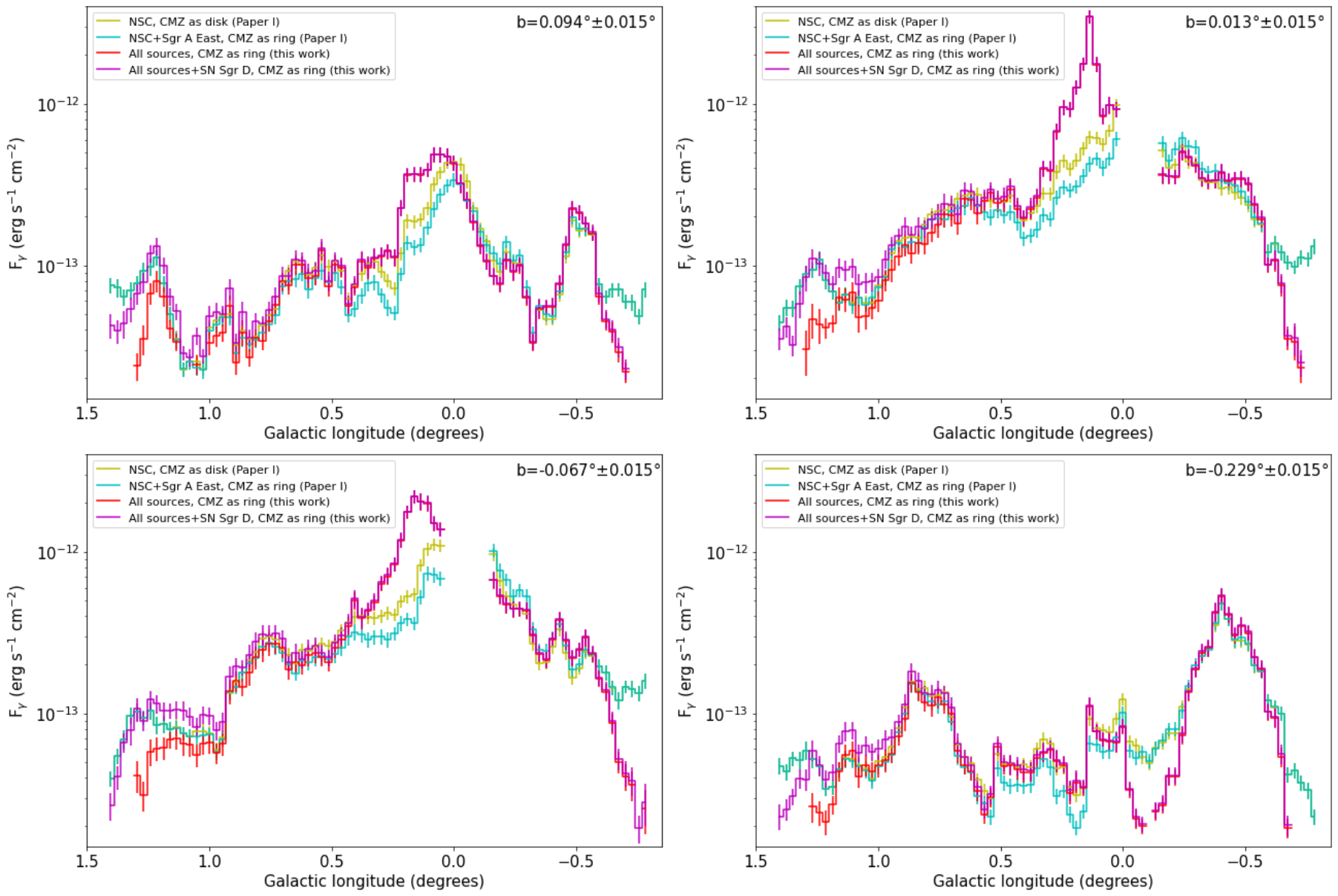}
\caption{Same as Fig.~\ref{fig_F-pro}, but now plotting all sources (NSC, Sgr A East, AC, and QC) plus SN Sgr D.}
\label{fig_Lon-D-appen}
\end{figure*}

\end{appendix}
\end{document}